\documentclass[a4paper,11pt]{article}%
\usepackage{amssymb,amsmath,amsfonts}
\usepackage{amsmath}
\usepackage{amsfonts}
\usepackage{amssymb}
\usepackage{graphicx}%
\providecommand{\U}[1]{\protect\rule{.1in}{.1in}}

\newtheorem{idea memo}[theorem]{Idea Memo}

\begin{document}

\title{Micro-Macro Duality and Emergence of Macroscopic Levels\thanks{Talk at the
International Symposium, QBIC 2007. }}
\author{Izumi OJIMA\\RIMS, Kyoto University}
\date{}
\maketitle

\begin{abstract}
The mutual relation between quantum Micro and classical Macro is clarified by
a unified formulation of instruments describing measurement processes and the
associated amplification processes, from which some perspective towards a
description of emergence processes of spacetime structure is suggested.

\end{abstract}

\section{Sectors as Quantum-Classical Boundary}

To fix the setting of my discussion, let me start from a brief account of the
notions of \textit{sectors}, \textit{inter-sectorial structures},
\textit{order parameters} to parametrize sectors and so on. In terms of these
we can formulate in a clear-cut manner the most important aspects of the
mutual relations\ between the microscopic quantum world and the macroscopic
classical levels, which is to be interpreted as a mathematical formulation
\cite{Unif03} of the physically essential idea of \textquotedblleft
quantum-classical correspondence\textquotedblright: the \textquotedblleft
boundary\textquotedblright\ and the gap between the former with
non-commutative algebras of quantum physical variables and the latter with
commutative algebras can be described by means of the notion of a
(\textit{superselection})\textit{\ sector structure }consisting of a family of
\textit{sectors }(or \textit{pure phases}). To define it, we need to classify
representations and states of a C*-algebra $\mathfrak{A}$ of quantum
observables according to the \textit{quasi-equivalence} $\pi_{1}%
\thickapprox\pi_{2}$ \cite{Dix} defined by the unitary equivalence of
representations $\pi_{1},\pi_{2}$ \textit{up to multiplicity}, which is
equivalent to the isomorphism of representing von Neumann algebras $\pi
_{1}(\mathfrak{A)}^{\prime\prime}\simeq\pi_{2}(\mathfrak{A)}^{\prime\prime}$.
A \textit{sector}, or, a \textit{pure phase} in the physical contexts, is then
defined by a quasi-equivalence class of \textit{factor} representations and
states corresponding to a von Neumann algebra with a trivial centre which is a
minimal unit among quasi-equivalence classes.

Representations belonging to different sectors $\pi_{a},\pi_{b}$ are mutually
\textit{disjoint\ }with no non-zero intertwiners, namely, any bounded operator
$T$ from the representation space $\mathfrak{H}_{\pi_{a}}$ of $\pi_{a}$ to
that $\mathfrak{H}_{\pi_{b}}$ of $\pi_{b}$ vanishes, $T=0$, if it satisfies
$T\pi_{a}(A)=\pi_{b}(A)T$ for $\forall A\in\mathfrak{A}$. If $\pi$ is not a
factor representation belonging to a sector, it can be uniquely decomposed
into the direct sum (or integral) of sectors, through the spectral
decomposition of a non-trivial commutative algebra $\mathfrak{Z}%
(\pi(\mathfrak{A)}^{\prime\prime})=\pi(\mathfrak{A})^{\prime\prime}\cap
\pi(\mathfrak{A})^{\prime}=\mathfrak{Z}_{\pi}(\mathfrak{A})$ as the centre of
$\pi(\mathfrak{A)}^{\prime\prime}$ admitting a \textquotedblleft simultaneous
diagonalization\textquotedblright. Here each sector contained in $\pi$\ is
faithfully parametrized by the Gel'fand spectrum $Spec(\mathfrak{Z}_{\pi
}(\mathfrak{A}))$ of the centre. Thus, commutative classical observables
belonging to the centre $\mathfrak{Z}_{\pi}(\mathfrak{A})$ physically plays
the role of \textit{macroscopic order parameters} and $Spec(\mathfrak{Z}_{\pi
}(\mathfrak{A}))$ can be regarded as the \textit{classifying space of sectors}
to distinguish different sectors. In this way, we find in a \textit{mixed
phase} the coexistence of quantum(=intra-sectorial) and classical systems,
which constitute an \textit{inter-sectorial} structure concisely described by
the centre $\mathfrak{Z}_{\pi}(\mathfrak{A})$ consisting of order parameters.

The traditional understanding of a sector is a \textquotedblleft coherent
subspace\textquotedblright\ where the \textquotedblleft superposition
principle\textquotedblright\ holds, but this \textquotedblleft
definition\textquotedblright\ applies only to sectors containing irreducible
representations and pure states which are meaningful only in the contexts
discussing the global aspects of quantum fields in the \textit{vacuum
}situation. Moreover, it leads to such a misleading interpretation of a
\textquotedblleft superselection rule\textquotedblright\ as an obstruction to
the superposition\ of state vectors belonging to different sectors; actually
the superposition of this sort is never \textquotedblleft
forbidden\textquotedblright\ but it simply reduces to statistical
\textit{mixtures} instead of superposed pure states, for lack of observables
with non-vanishing off-diagonal terms connecting different sectors. In sharp
contrast, the above general definition based on factoriality is applicable to
any pure phases associated with reducible factor representations and mixed
states which are common in the thermal and/or local aspects of quantum fields
(latter even in the vacuum situations), owing to the inevitable relevance of
\textit{non-type I representations} (for which irreducible representations are
almost useless).

\section{Instruments for Intra-sectorial Searches}

While the \textit{inter-sectorial }structure can successfully be treated by
means of the notions of sectors and of the macroscopic order parameters
belonging to the centre, this is not sufficient for a satisfactory description
of a given quantum system unless we combine it with the analysis of the
intrinsic quantum structures \textit{within} each sector, not only
theoretically but also operationally (up to the resolution limits imposed by
quantum theory itself). Since all the states in a sector share the same
spectrum of the centre, however, the order parameters are of little use in the
search of the \textit{intra-sectorial }structures within a sector. For the
purpose of detecting these invisible microscopic quantum structures we need a
general scheme of quantum measurement which has been proposed in \cite{Oji05,
OT06} by extending the standard scheme \cite{Oza} to systems with infinite
degrees of freedom. This is based upon the notion of a maximal abelian
subalgebra (MASA, for short)\ $\mathcal{A}$ of a factor von Neumann algebra
$\mathcal{M}=\pi(\mathfrak{A})^{\prime\prime}$ describing a fixed sector,
defined by the relation $\mathcal{A}=\mathcal{A}^{\prime}\cap\mathcal{M}$; if
we adopted the familiar condition $\mathcal{A}=\mathcal{A}^{\prime}$ it would
exclude the cases with $\mathcal{M}$ of non-type I common in quantum systems
with infinite degrees of freedom. Given such a MASA $\mathcal{A}%
=\mathcal{A}^{\prime}\cap\mathcal{M}$, the precise form of the
\textit{measurement coupling} can be specified between the observed system and
the apparatus required for implementing a measurement, on the basis of which
the central notion of \textit{instrument} can concisely be formulated. The
essence of the formulation can be summarized in terms of the following basic ingredients:

\begin{enumerate}
\item A (factor) von Neumann algebra $\mathcal{M}$(:$=\pi(\mathfrak{A}%
)^{\prime\prime}$) describing the observed system (in a fixed sector $\pi$)
and its MASA $\mathcal{A}=\mathcal{M}\cap\mathcal{A}^{\prime}=\mathcal{M}%
^{\mathcal{U}(\mathcal{A})}$ with the group $\mathcal{U}(\mathcal{A})$ of all
unitaries in $\mathcal{A}$. Under the physically natural assumption that the
representation Hilbert space $\mathfrak{H}_{\pi}$ of the present system
$\mathcal{M}$ can be taken as separable, $\mathcal{A}$ as observables to be
measured is generated by a locally compact abelian (Lie) group $\mathcal{U}%
\subset\mathcal{A}=\mathcal{U}^{\prime\prime}$ (with a Haar measure $d\mu$).
Since the results of a measurement of $\mathcal{A}$ are given by the measured
data belonging to $Spec(\mathcal{A})$, the \textit{algebra of the measuring
system\textbf{\ }}can be identified with the subalgebra $\mathcal{A}$ itself
of the observed system $\mathcal{M}$.

\item The \textit{measurement coupling} between the observed and the measuring
systems is specified by a\textit{\ }representation $U(W)$ of the
\textit{Kac-Takesaki operator} (K-T operator, for short)$\ W$ of the group
$\mathcal{U}$ defined by $(W\eta)(u,v):=\eta(v^{-1}u,v)$ for $\eta\in
L^{2}(\mathcal{U}\times\mathcal{U},d\mu\otimes d\mu),u,v\in\mathcal{U}$ and
characterized by the so-called pentagonal relation $W_{12}W_{23}=W_{23}%
W_{13}W_{12}$. When the action $\mathcal{M}\underset{\alpha}{\curvearrowleft
}\mathcal{U}$\ of the measuring system is implemented, $\alpha_{u}%
(M)=U_{u}MU_{u}^{-1}$ ($M\in\mathcal{M}$, $u\in\mathcal{U}$), by a unitary
representation $U$ of $\mathcal{U}$ on the (standard) representation Hilbert
space $L^{2}(\mathcal{M})$ of $\mathcal{M}$, the representation $U(W)$ of $W$
corresponding to $\alpha=AdU$ is defined by
\[
(U(W)\xi)(u):=U_{u}(\xi(u))\text{ \ \ \ for }\xi\in L^{2}(\mathcal{M})\otimes
L^{2}(\mathcal{U},d\mu),
\]
satisfying the (modified) pentagonal relation
\[
U(W)_{12}W_{23}=W_{23}U(W)_{13}U(W)_{12},
\]
and the intertwining relation $U(W)(1\otimes\lambda_{u})=(U_{u}\otimes
\lambda_{u})U(W)$. Here the suffices indicate the positions in the tensor
product $L^{2}(\mathcal{M})\otimes L^{2}(\mathcal{U})\otimes L^{2}%
(\mathcal{U})$ to which the operators act and $\lambda_{u}$ is the regular
representation of $\mathcal{U}$ defined by $(\lambda_{u}\eta)(v):=\eta
(u^{-1}v)$ on $\eta\in L^{2}(\mathcal{U})$. The simplest standard choice of
$\alpha$ common in the context of measurements is $\alpha_{u}(M)=uMu^{-1}$
(for $M\in\mathcal{M}$), $U_{u}=u$, which neglects the effect of the intrinsic
dynamics of the observed system on the measurement process. In terms of the
Lie generators $X_{a}$ of the unitary representation $U$ such that $U_{u}%
=\exp(\sum_{a}X_{a}\varphi^{a}(u))$, the coupling term can be written by
$U(W)=\exp(X_{a}\otimes\varphi^{a}(\hat{u}))$, where $\varphi^{a}(\hat{u})$
denotes an operator on $L^{2}(\mathcal{U})$ defined by $(\varphi^{a}(\hat
{u})\eta)(u)=\varphi^{a}(u)\eta(u)$\ for $\eta\in L^{2}(\mathcal{U}%
),u\in\mathcal{U}$ (similarly to the position operator $\hat{x}$ in quantum
mechanics, where the displacement unitary $\lambda_{x}=\exp(-i\hat{p}x)$
corresponds to the unitary operator $\lambda_{u}$ in the present context).

\item By restriction to $\mathcal{U}$ our measured data $\chi\in
Spec(\mathcal{A})$ can be embedded as a group character $\chi\upharpoonright
_{\mathcal{U}}$ of $\mathcal{U}$ into the dual group $\widehat{\mathcal{U}}$
which is again a locally compact abelian group. By Fourier-transforming $U(W)$
to $\widetilde{U(W)}:=(id\otimes\mathcal{F})U(W)(id\otimes\mathcal{F})^{-1}$
with $(\mathcal{F}\xi)(\gamma):=\int_{\mathcal{U}}\overline{\gamma(u)}%
\xi(u)d\mu(u)$ for $\xi\in L^{2}(\mathcal{U},d\mu)$, we define an
\textit{instrument} $\mathcal{I}$ for measuring $\mathcal{A}$ by
\[
\mathcal{I}(\Delta|\omega)(M):=(\omega\otimes m_{\mathcal{U}})\left(
\widetilde{U(W)}(M\otimes\chi_{\Delta})\widetilde{U(W)^{\ast}}\right)  ,
\]
for $M\in\mathcal{M}$, $\chi_{\Delta}\in\mathcal{A}=L^{\infty}%
(Spec(\mathcal{A}))$. While the identity element $\iota\in\widehat
{\mathcal{U}}$ for a non-compact $\mathcal{U}$ is not represented by a
normalized vector in $L^{2}(\mathcal{U})$, the above invariant mean
$m_{\mathcal{U}}$ over $\mathcal{U}$ physically plays the role of the
\textit{neutral position} $\iota$ of the measuring apparatus. All the
ingredients relevant to a measurement process are incorporated in this
instrument $\mathcal{I}$, such as the probability distribution $p(\Delta
|\omega)=\mathcal{I}(\Delta|\omega)(\mathbf{1})$ of measured values of
observables in $\mathcal{A}$ to be found in a Borel set $\Delta\subset
Spec(\mathcal{A})$ and as the state change from an initial state $\omega$ to a
final state $\mathcal{I}(\Delta|\omega)/p(\Delta|\omega)$ caused by the
read-out of measured values $\in\Delta$ \cite{Oza}, according to which a
process of the so-called \textquotedblleft reduction of wave
packets\textquotedblright\ is described.

\item Since $\mathcal{U}$ is abelian, we can consider the spectral
decomposition, $U_{u}=\int_{\chi\in Spec(\mathcal{A})\subset\widehat
{\mathcal{U}}}\overline{\chi(u)}dE(\chi)$ $(u\in\mathcal{U})$, of the unitary
representation $U$ (owing to the so-called SNAG theorem). Using this and the
Fourier transform $V=(\mathcal{F}\otimes\mathcal{F})W^{\ast}(\mathcal{F}%
\otimes\mathcal{F})^{-1}$ of $W$ as the K-T operator of the dual group
$\widehat{\mathcal{U}}$ with the Plancherel measure $d\hat{\mu}$ satisfying
the relation $(V\eta)(\gamma,\chi)=\eta(\gamma,\gamma^{-1}\chi)$\ for $\eta\in
L^{2}(\widehat{\mathcal{U}},d\hat{\mu})$, we have a clearer picture of
$\widetilde{U(W)}$: $\widetilde{U(W)}=\int_{\chi\in Spec(\mathcal{A})}%
dE(\chi)\otimes\lambda_{\chi}^{\ast}=:\widetilde{U}(V)^{\ast}$. In the Dirac
notation (of non-normalizable generalized eigenvectors), the action of
$\widetilde{U}(V)$ on $L^{2}(\mathcal{M})\otimes L^{2}(\widehat{\mathcal{U}})$
is given for $\gamma\in\widehat{\mathcal{U}}$, $\xi\in L^{2}(\mathcal{M})$ by
$\widetilde{U}(V)(\xi\otimes|\gamma\rangle)=\int_{\chi\in Spec(\mathcal{A}%
)}dE(\chi)\xi\otimes|\chi\gamma\rangle$.

\item In terms of the above K-T operators, the crossed product
$\mathcal{M\rtimes_{\alpha}U}$ is defined on $L^{2}(\mathcal{M})\otimes
L^{2}(\mathcal{U})$ as an important notion in the Fourier-Galois duality in
the following two equivalent ways: either as a von Neumann algebra
$\lambda^{\mathcal{M}}(L^{1}(\mathcal{U},\mathcal{M}))^{\prime\prime}$
generated by the Fourier transform $\lambda^{\mathcal{M}}(\hat{F}%
):=\int_{\mathcal{U}}\hat{F}(u)U(u)d\mu(u)$ of $\mathcal{M}$-valued $L^{1}%
$-functions$\ \hat{F}\in L^{1}(\mathcal{U},\mathcal{M})$ with the convolution
product, $(\hat{F}_{1}\ast\hat{F}_{2})(u)=\int_{\mathcal{U}}\hat{F}%
_{1}(v)\alpha_{v}(\hat{F}_{2}(v^{-1}u)d\mu(v)$, mapped by $\lambda
^{\mathcal{M}}$ into $\lambda^{\mathcal{M}}(\hat{F}_{1}\ast\hat{F}%
_{2})=\lambda^{\mathcal{M}}(\hat{F}_{1})\lambda^{\mathcal{M}}(\hat{F}_{2})$,
or, as a von Neumann algebra $\pi_{\alpha}(\mathcal{M})\vee(1\otimes
\lambda(\mathcal{U}))$ generated by $1\otimes\lambda(\mathcal{U})$ and by
\[
\pi_{\alpha}(\mathcal{M}):=\{\pi_{\alpha}(M):=Ad(U(W)^{\ast})(M\otimes
1);~M\in\mathcal{M}\}.
\]
These two versions are related by the mapping $\alpha(W):=Ad(U(W))$,
\[
\lambda^{\mathcal{M}}(L^{1}(\mathcal{U},\mathcal{M}))^{\prime\prime
}=(\mathcal{M}\otimes1)\vee\{U_{u}\otimes\lambda_{u};u\in\mathcal{U}%
\}\overset{\alpha(W)^{-1}}{\underset{\alpha(W)}{\rightleftarrows}}\pi_{\alpha
}(\mathcal{M})\vee(1\otimes\lambda(\mathcal{U}),
\]
which can be understood as the Schr\"{o}dinger and Heisenberg pictures: the
former $(\mathcal{M}\otimes1)\vee\{U_{u}\otimes\lambda_{u};u\in\mathcal{U}\}$
is in the Schr\"{o}dinger picture with \textit{unchanged }microscopic
observables $\mathcal{M}\otimes1$ and with the\textit{\ coupling}
$U_{u}\otimes\lambda_{u}$ \textit{to change macroscopic states}, while, in the
latter, all the coupling effects are concentrated in the observables
$\pi_{\alpha}(\mathcal{M})$ in contrast to the\textit{\ kinematical changes}
of macroscopic\textit{\ states} caused by $\lambda(\mathcal{U})$.
\newline\ \ \ In the case of the instrument $\mathcal{I}$, the algebra to be
observed is the tensor algebra $\mathcal{M}\otimes\mathcal{A}=\mathcal{M}%
\otimes L^{\infty}(Spec\mathcal{A})$ realized in the \textit{initial} and
\textit{final} stages, respectively, before and after the measuring processes
according to the switching-on and -off of the coupling $\alpha$:
$\mathcal{M}\otimes\mathcal{A}=\mathcal{M}\rtimes_{\alpha=id_{\mathcal{M}}%
}\mathcal{U}\rightarrow\mathcal{M}\rtimes_{\alpha}\mathcal{U}\rightarrow
\mathcal{M}\otimes\mathcal{A}$, similarly to the scattering processes. All the
effects of the measurement coupling $\widetilde{U(W)}$ are encoded in the form
of \textit{macroscopic state} \textit{changes} recorded in the spectrum of the
non-trivial centre $\mathfrak{Z}(\mathcal{M}\otimes\mathcal{A)}=\mathcal{A}%
=L^{\infty}(Spec\mathcal{A})$ of $\mathcal{M}\otimes\mathcal{A}$, playing the
same roles as the order parameters to specify sectors in the inter-sectorial
context. For these reasons, the most natural physical essence of the formalism
based on an instrument $\mathcal{I}$ should be found in the
\textit{interaction picture}, whose\textit{ coupling} term $\widetilde
{U(W)}=(id\otimes\mathcal{F})U(W)(id\otimes\mathcal{F})^{-1}$ is responsible
for deforming the decoupled algebra $\mathcal{M}\otimes\mathcal{A}$ into the
above crossed product $\mathcal{M}\rtimes_{\alpha}\mathcal{U}$.
\end{enumerate}

\section{Amplification in Intra-sectorial Measurements}

While the notion of an instrument provides a sufficient tool for the
operational description of a measurement, the above state changes\ describe
only \textit{microscopic} changes of quantum states $\xi\otimes|\iota
\rangle\rightarrow\xi_{\gamma}\otimes|\gamma\rangle$ of the composite system
of the observed system and the \textit{probe system} taking place at their
microscopic contact point. The question remains untouched as to how the
invisible microscopic changes in the quantum states are \textit{transformed
into visible macroscopic changes} of the measuring pointer, without which
measured values $\in Spec(\mathcal{A})\subset\widehat{\mathcal{U}}$ cannot be
read out or registered. To answer this question we need a mathematical
formulation of the process of \textit{amplification} from the microscopic
state changes in the probe system caused by the measurement coupling into the
macroscopic changes in the spatial positions of the measurement pointer. While
I am not aware of known results of this sort, this kind of amplifying
mechanism seems to be universally relevant to any bridges between
micro-quantum systems and macro-classical world.

In the present approach, the mathematical essence of the amplification
processes can be seen in the following simple form \cite{Oji06} based upon the
quasi-equivalence $\lambda^{\otimes m}\approx\lambda^{\otimes n}$ ($\forall
m,n\in\mathbb{N}$) among arbitrary tensor powers $\lambda^{\otimes n}%
=\lambda\otimes\cdots\otimes\lambda$~of the regular representation of a
locally compact group $\widehat{\mathcal{U}}$ via the K-T operator $V$ related
closely to the measurement coupling. When $V$ is applied arbitrarily many
times to an initial state $\xi\otimes|\iota\rangle\otimes|\iota\rangle
\cdots\otimes|\iota\rangle$\ of the composite system where $\xi=\sum
_{\gamma\in\widehat{\mathcal{U}}}c_{\gamma}\xi_{\gamma}$ is an initial state
of the observed system, the resulting state becomes:
\begin{align*}
&  V_{N,N+1}\cdots V_{23}\widetilde{U}(V)_{12}(\xi\otimes\underset
{N}{\underbrace{|\iota\rangle\otimes|\iota\rangle\cdots\otimes|\iota\rangle}%
})\\
&  =\sum_{\gamma\in\widehat{\mathcal{U}}}c_{\gamma}V_{N,N+1}\cdots
V_{34}V_{23}(\xi_{\gamma}\otimes|\gamma\rangle\otimes|\iota\rangle
\cdots\otimes|\iota\rangle)\\
&  =\sum_{\gamma\in\widehat{\mathcal{U}}}c_{\gamma}V_{N,N+1}\cdots V_{34}%
(\xi_{\gamma}\otimes|\gamma\rangle\otimes|\gamma\rangle\cdots\otimes
|\iota\rangle)=\cdots\\
&  =\sum_{\gamma\in\widehat{\mathcal{U}}}c_{\gamma}\xi_{\gamma}\otimes
\underset{N}{\underbrace{|\gamma\rangle\otimes|\gamma\rangle\cdots
\otimes|\gamma\rangle}}\text{ }\underset{N\gg1}{\rightarrow}\text{ }%
\sum_{\gamma\in\widehat{\mathcal{U}}}c_{\gamma}\xi_{\gamma}\otimes\left[
|\gamma\rangle^{\otimes N}\right]  ,
\end{align*}
(whose validity is, to be precise, restricted to the case with $\widehat
{\mathcal{U}}$ having a discrete spectrum). However, the corresponding formula
in the Heisenberg picture given by
\begin{align*}
&  A\otimes f_{2}\otimes\cdots\otimes f_{N+1}\\
&  \longmapsto\widetilde{U}(V)_{12}^{\ast}V_{23}^{\ast}\cdots V_{N,N+1}^{\ast
}(A\otimes f_{2}\otimes\cdots\otimes f_{N+1})V_{N,N+1}\cdots V_{23}%
\widetilde{U}(V)_{12}\\
&  =Ad(\widetilde{U}(V)_{12}^{\ast})Ad(V_{23}^{\ast})\cdots Ad(V_{V_{N,N+1}%
}^{\ast})(A\otimes f_{2}\otimes\cdots\otimes f_{N+1})\\
&  =Ad(\widetilde{U}(V))(A\otimes Ad(V^{\ast})(f_{2}\otimes Ad(V^{\ast
})(\cdots\otimes Ad(V^{\ast})(f_{N}\otimes f_{N+1})))\cdots)\\
&  \text{
\ \ \ \ \ \ \ \ \ \ \ \ \ \ \ \ \ \ \ \ \ \ \ \ \ \ \ \ \ \ \ \ \ \ \ \ for
}A\in\mathcal{M}\text{\ and }f_{i}\in L^{\infty}(\widehat{\mathcal{U}}),
\end{align*}
is similar to the one appearing in Accardi's formulation of quantum Markov
chain \cite{Acc74} which is indepedent of the discreteness of the spectrum.
According to the general basic idea of \textquotedblleft quantum-classical
correspondence\textquotedblright, a classical macroscopic object can be
identified with a condensed state of infinite number of quanta, as well
exemplified by the macroscopic magnetization of Ising or Heisenberg
ferromagnets described by the aligned states $|+\rangle^{\otimes\infty}$ of
infinite number of microscopic spins. Therefore, the above state
$|\gamma\rangle^{\otimes N}$ (with $N\gg1$) can physically be interpreted as
representing a macroscopic position of the measuring pointer, and hence, the
above repeated action of the K-T operator $V$ describes a cascade process or a
domino effect\ of \textquotedblleft decoherence\textquotedblright\ to amplify
a state change at the microscopic end of the apparatus into the macroscopic
classical motion $\iota\rightarrow\gamma$ of the measuring pointer. It is
remarkable here that the quasi-equivalence of arbitrary tensor powers
$\lambda^{\otimes n}$~of the regular representation $\lambda$ guarantees the
\textquotedblleft unitarity\textquotedblright\ of the above amplification
process, which provides the mathematical basis for not only the
\textquotedblleft repeatablity hypothesis\textquotedblright\ but also the
possibility of the recurrent quantum interference even after the contact with
the measuring apparatus under the situation that the number $N$ of repetition
need not be regarded as a real infinity (as the size of $N$ depends on the
length of the interval responsible for the amplification process between the
microscopic and macroscopic ends of the measurement apparatus and also on the
reaction rate of the flip from $|\iota\rangle$ to $|\gamma\rangle$). In this
respect, the problem as to whether the situation is completely made classical
or not depends highly on the relative configurations among many large or small
numbers, which can consistently be described in the framework of the
non-standard analysis (see, for instance, \cite{OjiOza}). In relation to this,
it is also interesting to note that the above amplification process is closely
related to a L\'{e}vy process through its \textquotedblleft infinite
divisibility\textquotedblright\ as follows: similarly to the affine property
$f(\lambda x+\mu y)=\lambda f(x)+\mu f(y)$ ($\forall\lambda,\mu>0$) of a map
$f$ defined on a convex set which follows from the addivitiy
$f(x+y)=f(x)+f(y)$, we can derive $\lambda\thickapprox\lambda^{n/m}$ ($\forall
m,n\in\mathbb{N}$) from $\lambda\thickapprox\lambda^{n}$ ($\forall
n\in\mathbb{N}$), which means the infinite divisibility\ $(AdV^{\ast}%
)^{t+s}\thickapprox(AdV^{\ast})^{t}(AdV^{\ast})^{s}$ ($t,s>0$) of the process
induced by the above transformation. In this way, we see that simple
individual measurements with definite measured values are connected without
gaps with discrete and/or continuous repetitions of measurements
\cite{OjiTan92}. If this formulation exhausts the essence of the problem, the
remaining tasks reduce to its physical and technological implementation
through suitable choices of the media connecting the microscopic contact point
between the system and the apparatus to the measuring pointer.

\section{From Amplification to Emergence of Macro}

In the mutual relations between invisible Micro and visible Macro, we find
interesting recurrent patterns among dynamical systems, crossed products to
formulate coupled systems and processes to amplify the results of state
changing processes into readable data. The crucial roles are played here by
the K-T operators and the Fourier duality to perform the spectral
decomposition. To understand their natural operational meaning we compare the
above scheme for an intra-sectorial search with the measurement of an
inter-sectorial structure associated with an unbroken internal symmetry, whose
basic ingredients are as follows:

\begin{enumerate}
\item A microscopic system described by a field algebra $\mathcal{\mathfrak{F}%
}$ and a (compact) group $G$ of internal symmetry constituting a dynamical
system $\mathcal{\mathfrak{F}}\underset{\alpha}{\curvearrowleft}G$.

\item The coupled system of observed and measuring systems is given by a
\textit{crossed product} $\mathfrak{F}\rtimes_{\alpha}G\simeq\mathfrak{F}%
^{G}\equiv\mathfrak{A}$ whose sector structure is parametrized by order
parameters belonging to the set $\widehat{G}$ of equivalence classes of
irreducible unitary representations of $G$.

\item Measured values (in a given representation $\pi$ of $\mathfrak{F}$) are
registered in $Spec(\mathfrak{Z}_{\pi}(\mathfrak{A}))=\widehat{G}$: note
the\textit{\textbf{\ }Fourier duality} between $G$ \textit{acting} on the
system and its dual $\widehat{G}$ as sector indices to be \textit{measured}.

\item The \textit{K-T operator} relevant to measured data in $\widehat{G}$ is
given in the form of $\hat{V}:=\sigma V^{\ast}\sigma$ defined in $L^{2}%
(\hat{G})=L^{2}(G)$ on the basis of the K-T operator $V$ of $G$ given by
$(V\xi)(g_{1},g_{2})=\xi(g_{1},g_{1}^{-1}g_{2})$ for $g_{1},g_{2}\in G$ (where
$\sigma$ is the flip operator on the tensor product Hilbert space). For an
abelian $G$, we have through the Fourier transform $(\hat{V}\eta)(\gamma
_{1},\gamma_{2})=\eta(\gamma_{1},\gamma_{1}^{-1}\gamma_{2})$ for $\gamma
_{1},\gamma_{2}\in\hat{G}$, which cannot, however, be literally reproduced for
a \textit{non-abelian} $G$ owing to the relevance of \textit{multi-dimensional
vector spaces} to representations $\gamma\in\hat{G}$ of $G$.
\end{enumerate}

In contrast, the problem of \textit{parameter estimate} in covariant
measurements is formulated as follows\U{ff1a}

\begin{enumerate}
\item an algebra to be observed is~$\mathcal{\mathfrak{A}}$ or $\mathfrak{F}%
\rtimes_{\alpha}G=\mathfrak{A}\otimes\mathcal{K}(L^{2}(G))$.

\item The coupling between $\mathcal{\mathfrak{A}}$ and $\hat{G}$\ due to the
\textit{co-action} $\mathcal{\mathfrak{A}}\curvearrowleft\hat{G}$\ leads to a
crossed product $\mathcal{\mathcal{\mathfrak{A}}\rtimes}\hat{G}\simeq
\mathfrak{F}$ as a measurement is a process to \textit{couple the system to
the dual variables of what to be observed}.

\item What to be read out in this case as the outcome of the measurement is
$g\in G$ whose non-commutativity requires an optimized choice of
\textit{positive operator-valued measures} (\textit{POVM's}, for
short)\textbf{\ }defined on $G$ taking values in the representation space of
$\mathfrak{F}$.

\item In the \textit{Naimark extension\textbf{\textit{\ }}}of a
POVM\textit{,\textbf{\textit{\ }}}the augmented algebra $\widehat
{\mathfrak{F}}$ of $\mathfrak{F}$ appears with a centre $\mathfrak{Z}%
(\widehat{\mathfrak{F}})=L^{\infty}(G)$ whose spectrum is $G$ (see
\cite{Unif03}).
\end{enumerate}

The duality of crossed products relevant to the above two cases can be
summarized as follows:
\begin{align*}
&  \left[
\begin{array}
[c]{cc}%
\begin{array}
[c]{c}%
\text{coupled system}\\
\mathcal{\mathfrak{F}}\rtimes_{\alpha}G\simeq\mathfrak{A}%
\end{array}
\text{ \ \ \ \ \ } &
\begin{array}
[c]{c}%
\\
\overset{\text{amplify}}{\underset{V^{\otimes}}{\Longrightarrow}}%
\end{array}%
\begin{array}
[c]{c}%
\text{read-out }\in\hat{G}\\
\leftrightarrow\text{sectors}%
\end{array}
\\%
\begin{array}
[c]{c}%
\text{\ \ \ \ \ }\Uparrow\text{\ \ }\\
G\curvearrowright
\end{array}
\text{\ \ \ }%
\begin{array}
[c]{c}%
\circlearrowright\hat{G}\text{: coaction}\\
\Downarrow\text{ \ \ \ \ \ \ \ \ \ \ \ \ \ \ }%
\end{array}
& \\
\mathcal{\mathfrak{F}}\simeq\mathcal{\mathfrak{A}}\rtimes_{\hat{\alpha}}%
\hat{G}\text{ \ \ \ \ \ \ } &
\end{array}
\right] \\
&  \overset{\text{dual}}{\rightleftarrows}\left[
\begin{array}
[c]{cc}%
\begin{array}
[c]{c}%
\text{coupled system}\\
\mathcal{\mathfrak{A}}\rtimes_{\hat{\alpha}}\hat{G}\simeq\mathfrak{F}%
\end{array}
\text{ \ \ \ \ \ } &
\begin{array}
[c]{c}%
\\
\overset{\text{amplify}}{\Longrightarrow}%
\end{array}%
\begin{array}
[c]{c}%
\text{read-out}\\
\in G
\end{array}
\\%
\begin{array}
[c]{c}%
\text{\ \ \ \ \ }\Uparrow\text{\ \ }\\
\hat{G}\curvearrowright
\end{array}
\text{\ \ \ \ \ }%
\begin{array}
[c]{c}%
\circlearrowright G\text{: action}\\
\text{ }\Downarrow\text{\ \ \ \ \ \ \ \ \ \ }%
\end{array}
& \\
\mathcal{\mathfrak{A\simeq F}}\rtimes_{\alpha}G\text{ \ \ \ \ \ \ } &
\end{array}
\right]
\end{align*}

We encounter a physically natural context of this sort in the description of
the sector structure associated with\textit{\textbf{\ }}a \textit{spontaneous
symmetry breakdown (SSB)} of a bigger\textit{\textbf{\ }}goup $G$ into an
unbroken symmetry with its closed subgroup $H$, as follows:

\begin{enumerate}
\item The inter-sectorial structure (I) consisting of degenerate
\textquotedblleft vacua\textquotedblright\ associated with SSB: the breakdown
of an internal symmetry described by a group $G$ is known to cause the
violation of the Haag duality $\mathfrak{A}(\mathcal{O}^{\prime}%
)=\mathfrak{A}(\mathcal{O})^{\prime}$ for the starting local net
$\mathcal{O}\longmapsto\mathfrak{A}(\mathcal{O})=\mathfrak{F}(\mathcal{O}%
)^{G}$ of observable elements of quantum fields. Then it can be extended to
the Haag dual net given by $\mathfrak{A}^{d}(\mathcal{O}):=\mathfrak{A}%
(\mathcal{O}^{\prime})^{\prime}$ to recover the Haag duality. Through the
Doplicher-Roberts reconstruction \cite{DHR} applied to $\mathfrak{A}^{d}$, we
find a field algebra $\mathfrak{F}=\mathfrak{A}^{d}\rtimes\hat{H}$ with a
compact Lie group $H$ as a subgroup of $G$ to describe an unbroken symmetry of
$\mathfrak{F}$. Using the method developed in \cite{Unif03}, we can construct
an \textit{augmented algebra} $\widehat{\mathfrak{F}}=\mathfrak{A}^{d}%
\rtimes\hat{G}=\mathfrak{F}\rtimes\widehat{(H\backslash G)}$ from the
co-action of $G$ on $\mathfrak{A}^{d}$ or equivalently from that of a
homogeneous space $H\backslash G$ on $\mathfrak{F}$ such that its induced
representation $\hat{\pi}$ from the vacuum representation of $\mathfrak{F}$
has automatically the unitary implementers of the broken $G$ and that it has a
non-trivial centre $L^{\infty}(G/H)=L^{\infty}(G)^{H}$ on which the action of
$G$ is ergodic. In this way, the degenerate \textquotedblleft
vacua\textquotedblright\ consisting of the base space $G/H$ of the bundle of
sectors can be detected as the spectrum of the order parameter $\mathfrak{Z}%
_{\hat{\pi}}(\widehat{\mathfrak{F}})=L^{\infty}(G/H)$. The above second case
of the parameter estimate of $G$ in covariant measurements in the use of a
POVM can be reproduced if we take $H=\{e\}$ here. Note the parallelism between
the dynamical system $G\curvearrowright G/H$ and the \textit{Galois group}
$G$\textit{ }in \textit{classical Galois theory} acting on the \textit{space}
$G/H$\textit{\ of} \textit{solutions}.

\item The inter-sectorial structure (II) concerning sectors arising from the
unbroken symmetry $H$ on one of degenerate \textquotedblleft
vacua\textquotedblright: the above Haag dual net algebra $\mathfrak{A}%
^{d}=\mathfrak{F}^{H}$ can be regarded as a coupled system $\mathfrak{F}%
\rtimes H\simeq\mathfrak{F}^{H}=\mathfrak{A}^{d}$ of the field algebra
$\mathfrak{F}$ with its unbroken symmetry group $H$ arising from the action of
$H$: $\mathfrak{F}\curvearrowleft H\Longrightarrow\mathfrak{F}\rtimes
H=\mathfrak{F}^{H}=\mathfrak{A}^{d}$ in the use of the Takesaki-Takai duality
of crossed product. This coupled system is acted on by the group dual
$\widehat{H}$, the latter of which can be measured to describe the sector
structure of the unbroken $H$ on a \textquotedblleft vacuum\textquotedblright%
\ chosen among degenerate \textquotedblleft vacua\textquotedblright\ (by means
of, e.g., Casimir operators of $Lie(H)$). In this way, the sector structure
due to a spontaneously broken symmetry constitutes a sector bundle
$G\times_{H}\hat{H}\twoheadrightarrow G/H$ over the homogeneous space $G/H$
with a standard fibre $\hat{H}$.

\item Intra-sectorial structure: detected by means of a suitable MASA
(corresponding to a Cartan subalgebra of $Lie(H)$, for instance) of a factor
algebra $\pi_{\eta}(\mathfrak{F}^{H})^{\prime\prime}=\pi_{\eta}(\mathfrak{A}%
^{d})^{\prime\prime}$.
\end{enumerate}

The relation above among a POVM of the space $G/H$, its Naimark extension and
the augmented algebra $\widehat{\mathfrak{F}}$ with $\mathfrak{Z}_{\hat{\pi}%
}(\widehat{\mathfrak{F}})=L^{\infty}(G/H)=\mathfrak{Z}_{\hat{\pi}%
}(\mathfrak{A}^{d})$ can be naturally understood by means of the Stinespring
theorem of dilations based upon the complete positivity of a POVM. Note here
the mutual relations among \textit{condensates, Goldstone modes and domain
structures}: in SSB with $G$ broken down to $H$, the \textit{condensates
}and\textit{\ Goldstone modes} are both related to $G/H$ but in quite a
different manner. In the case with a Lie group $G$ describing the
spontaneously broken symmetry, the former corresponds to the base space $G/H$
of the tangent bundle $T(G/H)$ and the latter to the fibre space $T_{\dot{g}%
}(G/H)$ at each point\textit{\textbf{\ }}$\dot{g}\in G/H$ as follows:

\begin{enumerate}
\item Condensates (responsible for SSB): the list of all the possible
condensates can be so parametrized by $G/H$ that \textit{each sector
corresponds to a point }$\dot{g}\in G/H$. I.e., the relation of $G/H$ to the
condensates is that the set $G/H$ exhausts all the possible choices of
degenerate \textquotedblleft vacua\textquotedblright, among which only one
point of $G/H$ can be realized as a sector at each time.

\item Goldstone modes describe \textit{virtual fluctuations around a fixed
choice among the above condensates} without changing it.

\item In the case with \textit{phase coexistence}, different choices of the
condensates are realized in different regions of the real space through which
a domain-structure is realized. \textquotedblleft Real space\textquotedblright%
\ may be misunderstood as prior to the emergence of different phases, whereas
such a \textquotedblleft real space\textquotedblright\ may not be materialized
without the coexistence of phases.
\end{enumerate}

This last remark will play crucial roles in understanding classical
geometrical structures visible at the macroscopic levels as something arising
from the processes of \textit{emergence} from the invisible microscopic worlds.

\vskip12pt Last but not least, I would like to express my sincere thanks to
Prof. M. Ohya and Prof. N. Watanabe for the invitation to this pioneering and
inspiring International Conference QBIC2007 and to Prof. L. Accardi and to
Prof. T. Hida for their encouragements.

\end{document}